# New Threats to SMS-Assisted Mobile Internet Services from 4G LTE:
Lessons Learnt from Distributed Mobile-Initiated Attacks Towards Facebook and Other Services

Guan-Hua Tu[†], Yuanjie Li[†], Chunyi Peng[‡], Chi-Yu Li[†], Muhammad Taqi Razay[†], Hsiao-Yun Tseng[†], Songwu Lu[†]

[†] Department of Computer Science, University of California, Los Angeles
[‡] Department of Computer Science Engineering, The Ohio State University

*Abstract*—Mobile Internet is becoming the norm. With more personalized mobile devices in hand, many services choose to offer alternative, usually more convenient, approaches to authenticating and delivering the content between mobile users and service providers. One main option is to use mobile text service (*i.e.*, short messaging service). Such carrier-grade text service has been widely used to assist versatile mobile services, including social networking, banking, e-commerce, mobile-health, to name a few. Though the text service can be spoofed via certain Internet text service providers which cooperated with carriers, such attacks haven well studied and defended by industry due to the efforts of research community.

However, as cellular network technology *independently* advances to the latest IP-based 4G LTE, we find that these mobile services are somehow exposed to new threats raised by this change, particularly on 4G LTE Text service (via brand-new distributed Mobile-Initiated Spoofed SMS attack which is not available in legacy 2G/3G systems). The reason is that messaging service over LTE has to shift from the a conventional circuit-switched (CS) design to the packet-switched (PS) paradigm as 4G LTE supports PS only. Due to this fundamental change, 4G LTE Text Service becomes inevitably open and possible to access. However, its shields to messaging integrity and user authentication are not in place, while the legacy security mechanisms deemed effective for 2G/3G text are in vain. As a consequence, such weaknesses can be exploited to launch attacks against a targeted individual, a large scale of mobile users and even service providers, from mobile devices. Current defenses against Internet-Initiated Spoofed SMS attacks cannot work well against the unprecedented attack. Our study shows that 53 of 64 mobile services over 27 industries are vulnerable to at least one above threat, and some are even prone to million-dollar loss[1]. We validate these proof-of-concept attacks in one major US carrier which supports more than 100 million users. We finally propose quick fixes and recommended remedies and discuss security insights and lessons we have learnt.

## I. INTRODUCTION

Mobile service[2] is gaining increasing popularity in recent years, thanks to the rapid deployment of 3G/4G mobile networks. For most of service providers, how to authenticate the content of messages exchanged between them and mobile users and them at the reasonable cost is very important. It shall satisfy three properties: (1) *content integrity*: the data is not manipulated by the man in the middle; (2) *origin authenticity*: the data comes from the identified mobile user, and (3) *non-repudiation of origin (accountability)*: the user cannot deny the origin of the data delivery. Among many options (*e.g.*, message authentication code, digital signature or bind service-level user session with the secure communication channel), the mobile text service[3] is one popular solution to deliver the messages and authenticate the content in practice. Our studies show that this approach has been broadly adopted by the diversified and famous mobile service providers including social networking (Facebook, Twitter), grocery (Walmart, Target), airline (American Airline), bank (Chase, Citi), apparel (A&F), courier (Fedex, UPS), and name a few (more examples are given in Table I).

The prosperity of the mobile text service based approach stems from three reasons. First, the text service well satisfies three desired properties. Each mobile user has a physical SIM card which store a cellular-specific private key. The delivery of text message within carrier networks is confidentially and integrally protected by using the keys derived from the private key. Though it has some security issues (unauthorized text messages sent by phone-side malware or spoofed text messages from Internet), thanks to efforts of research communities and industry [10], [12], [24], [25], [31], [47], [48], [52], they are well addressed in the legacy 2G/3G networks, at least for top four largest US carriers. Second, the text service is one of the fundamental services in mobile networks. It is available in almost every mobile phone. For service providers, it is a convenient way to reach 6 billion mobile users on the earth.

However, the 4G LTE network adopts a design paradigm shift for its text service. Different from 2G/3G, LTE is an all IP network. Its text service thus migrates from the traditional circuit-switched (CS) to the full packet-switched (PS) design. The natural question is: *Given the dramatic design change in the 4G text service, are the mobile services via text still safe as they are using 2G/3G text?*

Unfortunately, our study yields the negative answer. For carriers, authentication/confidentiality/integrity protections, similar to those in circuit-switched 2G/3G text messages, are not always applied. The mobile device and mobile service providers have not been updated with new security components in accordance with the carrier infrastructure upgrade. The ser-vice providers seem to make security as second-class citizens so that security semantics behind the usage interface may not upgrade and ensured as the technology evolves. The service interface for normal usage remains invariant but the underlying security semantics changes as the technology evolves.

---
[1] This work is done before August 15, 2015. All the threats and attacks described in the paper were validated before that.

[2] In this paper, we denote the mobile service to be the service is offered to the mobile phones.

[3] In this work, we will interchangeably use "text" and SMS (short message service) for a slight abuse of wording definition.

More specifically, our results show that 53 of 64 mobile services (Table I) will suffer from new attacks, caused by the change of underlying security semantics. Due to space limit and similarity, we do not list all studied mobile services including USPS, Dollar Tree, ZipCar, Weather.com, Dick's Sporting *etc.*. Only 9 are free from the three attacks. We are able to (1) abuse the individual or large-scale Facebook[4] account(s) to update their status, like pages or add friend, (2) launch *donation attack* to transfer victims' money and (3) *malicious spam lawsuit attack* towards the brand-name companies. Our assumed attack model is simple: the attacker only uses commodity smartphones and have no control to carrier networks. Note that we do not require to install any malware on the victims' phones. Though, to avoid the tracing back issue, attacker may distribute a unprivileged application (malware) to other mobile users and turn them into the Bots to launch attacks. We evaluate these attacks in a responsive and controlled manner (*i.e.*, only attacking the participants of this project) in one major US carrier which supports more than 100 million users.

These attacks are the result of security vulnerabilities from the mobile device, carriers, mobile network standards and the mobile service providers. For mobile devices, the phone-side security mechanisms and permission control remain invariant while text service technology evolved, as well as they are easy to be bypass. For carriers, their security mechanisms do not well recognize the fundamental differences between 2G/3G and 4G text messages. 4G users can freely specify the originating address of the text message being sent which it is not allowed in 2G/3G by the protocol design. For mobile network standards (3GPP and 3GPP2[5]), they have a conflict that which the security mechanisms should be used for sending/receiving text message in 4G LTE. For mobile service providers, their loose service enrollment and weak (or even no) runtime authentication mechanisms make them vulnerable to the new attacks launched from 4G LTE mobile devices. We propose solutions to these vulnerabilities, and are working with all parties involved for the fix.

The rest of this paper is structured as follows. §II introduces the content authentication mechanisms used by mobile services, and how they cooperate with the legacy 2G/3G text service, and reviews the early security threats and defenses. §III describes the 4G text service, potential security vulnerabilities, attack model and methodology. §IV uses Facebook as a case study to illustrate the vulnerabilities exposed to mobile service users. §V presents the way to scale the individual Facebook attack to large-scale attack. §VI conducts a medium-scale study on mobile services, presents their security loopholes in common, and what possible attacks may be launched them. §VII proposes the solutions, §VIII describes the related work

---

[4]This attack is also applied to Twitter, as well.
[5]3GPP stipulates the standards of GSM, GPRS, WCDMA and UMTS (the representative US carriers are AT&T and T-Mobile), while 3GPP2 defines CDMAOne, CDMA200, EV-DO standards (the representative US carriers are Verizon and Sprint).

and §IX concludes the paper.

## II. BACKGROUND

### A. Content Authentication in Mobile Services

For mobile services, it is a critical function to authenticate the content of messages exchanged between mobile users and mobile service providers (*e.g.* social networking, banking transfer, donating). They include sensitive information, (user identity, money being transferred, *etc.*), and should be protected at runtime to guarantee three properties: (1) *content integrity*: the data is not manipulated by the man in the middle; (2) *origin authenticity*: the data comes from the identified mobile user, and (3) *non-repudiation of origin (accountability)*: the user cannot deny the origin of the data delivery.

In practice, there are three popular approaches to achieve the three goals. In the first approach, each delivered message is associated with a self-certifying fingerprints to enforce the security. This fingerprints can be either based on the lightweight shared secret between user and server (*e.g.* Message Authentication Code or MAC), or the heavyweight asymmetric public/private keys distributed to suer and server (*e.g.* digital signature). This approach can guarantee the integrity and authenticity of the message. However, the lightweight shared-secret cannot provide strong guarantee of non-repudiation (both users and service providers share the same secret), while the heavyweight asymmetric cryptography requires both users and service servers to have private keys issued by authorized CA, which is proven to be not scalable.

In the second approach, the content identity is bound to the communication channel between mobile user and service provider. For example, many mobile services bind username/password with the SSL/TLS connection between user and server. If the verification of username/password succeeds, the service server will generate a security token (*e.g.* HTTP cookies) associated with SSL/TLS connection between user and server. Similar to the first approach, the binding approach also guarantees the integrity of the messages. But the authenticity of the messages is usually unidirectional. That is, by SLL/TLS design, the messages from servers are always guaranteed with authenticity. But the authenticity of the messages from the user relies on the security token plus SSL/TLS connection context, which cannot provide strong non-repudiation property (users do not have unique private keys issued by authorized CA).

The third approach is to rely on the text service in mobile network. They take advantage of readily deployed mobile text service security to guarantee above properties in a cost-effective way. In fact, the mobile text service is more secure than people expected. We consider the scenario that the mobile service provider communicates with its users by using a mobile phone to send/receive text messages within the same carrier network. Other scenarios will be discussed later. Before mobile phone sends/receivs texts, it has to perform the registration procedure towards mobile network by using the private key (similar to the private certificate issued by CA, but it is cellular specific) stored in the SIM card (a physical artifact). During registration procedure, the unique integrity and confidentiality keys are derived from the private key to encrypt/decrypt the control-plane signaling messages in mobile

| No. | Provider | Industry | Short Code | Enrollment | Service | Operations via Text | M | A | W | T | Threats |
|---|---|---|---|---|---|---|---|---|---|---|---|
| 1 | Walmart | Grocery | 63257 | One-step | Sub-Notif | Special offer, flyer, update | No | No | No | Yes | Spam lawsuit |
| 10 | CVS Pharmacy | Pharmacy | 35437 | One-step | Sub-Notif | Special offer, flyer, update | No | No | No | Yes | Spam lawsuit |
| 18 | Costco | Grocery | 71034 | 3-Step-Simple | Sub-Notif | Special offer, flyer, update | No | No | No | Yes | Spam lawsuit |
| 21 | JP Morgan Chase | Bank | 24273 | 4-Step-AuthCode | Req-Resp | Query balance, history, *etc.* | No | No | Yes | No | No |
| 23 | Bank of America | Bank | 692632 | 4-Step-Simple | Req-Resp | Query balance, history, *etc.* | No | No | Yes | No | No |
| 28 | Citi Bank | Bank | 692484 | 4-Step-AuthCode | Req-Resp | Query balance, history, *etc.* | No | No | Yes | No | No |
| 30 | Wells Fargo | Bank | 93557 | Four-Step | Req-Resp | Query balance, history, transfer money and *etc.* | Yes | No | Yes | No | No |
| 33 | HomeDepot | Retailing | 38688 | 3-Step-Simple | Sub-Notif | Special offer, flyer, update | No | No | Yes | No | Spam lawsuit |
| 36 | Target Baby | Grocery | 827438 | One-step | Sub-Notif | Special offer, flyer, update | No | No | Yes | No | Spam lawsuit |
| 36 | Target Store | Grocery | 827438 | 3-Step-Simple | Sub-Notif | Special offer, flyer, update | No | No | Yes | No | Spam lawsuit |
| 41 | State Farm | Insurance | 78836 | 3-Step-Simple | Sub-Notif | Text alerts | No | No | No | No | No∗ |
| 47 | UPS | Courier | 69877 | 3-Step-Simple | Sub-Notif | Order status | No | No | Yes | No | Spam lawsuit |
| 50 | Lowes | Retailing | 656937 | 3-Step-Simple | Sub-Notif | Special offer, flyer, update | No | No | No | Yes | Spam lawsuit |
| 65 | Fedex | Courier | 48773 | One-step | Sub-Notif | Order status | No | No | No | Yes | Spam lawsuit |
| 70 | American Airline | Airline | 35922 | One-step | Sub-Notif | Query Flight status | No | No | No | Yes | Spam lawsuit |
| 84 | Safeway | Grocery | 25374 | 3-Step-Simple | Sub-Notif | Special offer, flyer, update | No | No | No | No | No∗ |
| 88 | American Express | Bank | 692639 | 4-Step-Simple | Req-Resp | Query balance, history, *etc.* | No | No | Yes | No | No |
| 104 | TimeWarner Cable | ISP | 789789 | One-step | Sub-Notif | Special offer, flyer, update | No | No | No | Yes | Spam lawsuit |
| 105 | Macy | Store | 62442 | One-step | Sub-Notif | Special offer, flyer, update | No | No | No | Yes | Spam lawsuit |
| 134 | Staple | Grocery | 555444 | One-step | Sub-Notif | Special offer, flyer, update | No | No | No | Yes | Spam lawsuit |
| 138 | US Bank | Bank | 872265 | Two-Step | Req-Resp | Query balance, history, *etc.* | No | No | Yes | No | No |
| 157 | KOHL's | Grocery | 56457 | One-step | Sub-Notif | Special offer, flyer, update | No | No | No | Yes | Spam lawsuit |
| 161 | SouthWest Airline | Airline | 72743 | One-step | Sub-Notif | Flight status update | No | No | Yes | No | Spam lawsuit |
| 187 | Starbucks | Retailing | 22122 | One-step | Sub-Notif | Special offer, flyer, update | No | No | No | Yes | Spam lawsuit |
| 194 | Office Depot | Grocery | 33768 | One-step | Sub-Notif | Special offer, flyer, update | No | No | Yes | No | Spam lawsuit |
| 221 | Marriot | Hotel | 58682 | One-step | Sub-Notif | Confirmation | No | No | Yes | No | No |
| 242 | **Facebook** | Social Networks | 32665 | 4-Step-AuthCode | Req-Resp, Sub-Notif | Update status, add friends, like pages, (un)subscribe, *etc.* | No | No | Yes | No | **Account abuse** |
| 245 | Toys R US | Toy | 78697 | 3-Step-Simple | Sub-Notif | Special offer, flyer, update | No | No | No | Yes | Spam lawsuit |
| 250 | JC Penny | Store | 527365 | 3-Step-Simple | Sub-Notif | Special offer, flyer, update | No | No | No | Yes | Spam lawsuit |
| 260 | Bed Bath Beyond | Grocery | 239663 | One-step | Sub-Notif | Special offer, flyer, update | No | No | No | Yes | Spam lawsuit |
| 303 | Discover | Bank | 347268 | Two-Step | Req-Resp | Sub-Notif, Alert | No | No | Yes | No | No |
| 541 | Levis | Apparel | 84483 | One-step | Sub-Notif | Special offer, flyer, update | No | No | No | Yes | Spam lawsuit |
| 648 | Abercrombie & Fitch | Apparel | 231892 | 3-Step-Simple | Sub-Notif | Special offer, flyer, update | No | No | No | Yes | Spam lawsuit |
| 648 | Abercrombie kids | Apparel | 34824 | 3-Step-Simple | Sub-Notif | Special offer, flyer, update | No | No | No | Yes | Spam lawsuit |
| 648 | Hollister Co | Apparel | 743722 | 3-Step-Simple | Sub-Notif | Special offer, flyer, update | No | No | No | Yes | Spam lawsuit |
| NA | **Twitter** | Social Networks | 40404 | 4-Step-AuthCode | Req-Resp, Sub-Notif | Update status/profile, block, send text,(un)follow, *etc.* | No | No | Yes | No | **Account abuse** |
| NA | Domino Pizza | Fast Food | 366466 | 3-Step-Simple | Sub-Notif | Special offer, flyer, update | No | No | Yes | No | Spam lawsuit |
| NA | Paypal | ePayment | 729725 | 4-Step-AuthCode | Req-Resp | Query balance, history, *etc.* | No | No | Yes | No | No |
| NA | **Papa John** | FastFood | 47272 | One-step | Sub-Notif | Special offer, flyer, update | No | No | No | Yes | **Spam lawsuit** |
| NA | Charlotte Russe | Apparel | 78953 | One-step | Sub-Notif | Special offer, flyer, update | No | No | No | Yes | Spam lawsuit |
| NA | Payless Shoes | Apparel | 747474 | 3-Step-Simple | Sub-Notif | Special offer, flyer, update | No | No | No | Yes | Spam lawsuit |
| NA | Southern Class Clothing | Apparel | 313131 | One-step | Sub-Notif | Special offer, flyer, update | No | No | Yes | No | Spam lawsuit |
| NA | ULTA Beauty | Cosmetic | 80565 | One-step | Sub-Notif | Special offer, flyer, update | No | No | No | Yes | Spam lawsuit |
| NA | AMC Theaters | Entertainment | 242424 | One-step | Sub-Notif | Special offer, flyer, update | No | No | No | Yes | Spam lawsuit |
| NA | Kay Jewelers | Jewelry | 553935 | 4-Step-Simple | Sub-Notif | Special offer, flyer, update | No | No | No | Yes | Spam lawsuit |
| NA | NBC NEWS | Media | 67622 | One-step | Sub-Notif | News update | No | No | No | Yes | Spam lawsuit |
| NA | NEWS 9 | Media | 79640 | One-step | Sub-Notif | News update | No | No | No | Yes | Spam lawsuit |
| NA | TMZ Store | Media | 52966 | 3-Step-Simple | Sub-Notif | News update | No | No | Yes | No | Spam lawsuit |
| NA | WISH TV | Media | 36729 | One-step | Sub-Notif | News/Weather update | No | No | No | Yes | Spam lawsuit |
| NA | Cabelas | Sport | 247365 | 3-Step-Simple | Sub-Notif | Special offer, flyer, update | No | No | No | Yes | Spam lawsuit |
| NA | **Red Cross** | HumanAid | 90999 | Always on | Req-Resp | Donation | Yes | Weak | No | No$^a$ | **Donation** |
| NA | Arvada Council | Art | 20222 | Always on | Req-Resp | Donation | Yes | Weak | No | No$^a$ | Donation |
| NA | PhoenixArtMuseum | Art | 20222 | Always on | Req-Resp | Donation | Yes | Weak | No | No$^a$ | Donation |
| NA | American Library Association | Education | 41518 | Always on | Req-Resp | Donation | Yes | Weak | No | No$^a$ | Donation |
| NA | Claflin Univ. | Education | 52000 | Always on | Req-Resp | Donation | Yes | Weak | No | No$^a$ | Donation |
| NA | Robert Morris Univ. | Education | 501501 | Always on | Req-Resp | Donation | Yes | Weak | No | No$^a$ | Donation |
| NA | In-N-Out Burger | FastFood | 20222 | Always on | Req-Resp | Donation | Yes | Weak | No | No$^a$ | Donation |
| NA | Greater Boston Food Bank | FoodAid | 20222 | Always on | Req-Resp | Donation | Yes | Weak | No | No$^a$ | Donation |
| NA | Zindagi Trust | FoodAid | 80088 | Always on | Req-Resp | Donation | Yes | Weak | No | No$^a$ | Donation |
| NA | WorldFoodProgram | FoodAid | 50555 | Always on | Req-Resp | Donation | Yes | Weak | No | No$^a$ | Donation |
| NA | Los Angeles Police Foundation | Government | 20222 | Always on | Req-Resp | Donation | Yes | Weak | No | No$^a$ | Donation |
| NA | Aid for Aids | Medical | 41010 | Always on | Req-Resp | Donation | Yes | Weak | No | No$^a$ | Donation |
| NA | American Cancer Society | Medical | 41518 | Always on | Req-Resp | Donation | Yes | Weak | No | No$^a$ | Donation |
| NA | National Wildlife Federation | Nature | 25383 | Always on | Req-Resp | Donation | Yes | Weak | No | No$^a$ | Donation |
| *Total* | 64 | 27 | | 6 | 2 | | | | | | **53/64** |

Table I. SUMMARY OF 64 MOBILE SERVICES WITH TEXT ENABLED. NO. IS SHORT FOR FORTUNE 500 RANKING [5] (N/A: NON-PROFIT ORGANIZATIONS OR NOT IN TOP 1000 LARGEST COMPANIES). THE MAWT COLUMNS ARE M: MONEY INVOLVEMENT, A: RUNTIME AUTHENTICATION, W: ENROLLMENT VIA WEB, T: ENROLLMENT VIA TEXT. NOTE THAT DIFFERENT MOBILE SERVICES MAY SHARE THE SAME SHORT CODE. THE TEXT MESSAGE IS DISPATCHED ACCORDINGLY BY THE TEXT CONTENT.

$^a$The mobile service is automatically enrolled by carriers.

networks. Since text message is delivered over control-plane, it is confidentially and integrally protected by design. Thus, the message integrity, authenticity and non-repudiation properties are satisfied. Compared with the traditional approaches, the text-based approach does not require extra private certificate to support strong non-repudiation property since each mobile user has a physical SIM card store a cellular-specific private key.

Beside, text service is ubiquitous to every mobile phone, making it readily deployable for mobile services without extra infrastructure or software deployment. Those characteristics make it as a popular mechanism to deliver and authenticate the messages exchanged between mobile users and service providers (as shown in Table I). Moreover, it also help mobile service providers to better verify user identifier (prevent malicious users to lots of fake accounts) than other alternatives (e.g., verify by email or voice call). Many mobile service providers, e.g., WhatsApp, Viber, eBuddy XMS, Tango, Voypi, Naver LINE, use text message to verify user. Although it still has some security issues (e.g., spoofed text), they are known to the academia and carriers, and well addressed (at least in top largest four US carriers). Due to its popularity and broad impact, we next introduce how mobile service is cooperated with 2G/3G text services.

### B. Mobile Service via 2G/3G Text Service

We first review how mobile services are achieved via 2G/3G text service, and then highlight the new changes with 4G message service in next section. Figure 1 shows messaging service in 2G/3G. In 2G/3G network, the text message is exchanged in the circuit-switched (CS) domain which is mandated by Mobile Switching Center (MSC). The text message is treated as a cellular signaling message, rather than normal data. The sender encapsulates the text message in a uplink transport signaling message, and delivers it to the MSC . The MSC decapsulates the message, and relays the text message to the Short Message Service (SMS) servers. The SMS server provides the store-and-forward feature, and tries to forward the message to the receiver through the circuit-switched path. Once the message is successfully delivered, the SMSC would acknowledge to the sender with a delivery report via the signaling channel.

Most mobile services through 2G/3G text service follow similar reference models for deployment, as shown in Figure 1. Most mobile service providers communicate with mobile users with unique short codes (*e.g.* 32655 for Facebook), which are audited and granted by Common Short Code Administration (CSCA) in US. More specifically, it works as follows. When mobile user sends a text message to a short code, the text message would then be forwarded by carriers to the authorized aggregators, which would further relay the message to the service provider. Note that the aggregators participating in the short code services have to be audited and monitored by CSCA and CITA (Cellular Telecommunications Industry Association). The service provider would then provide the corresponding service to user (*e.g.* subscribe, money transfer), and may reply the user with a text message.

**Threats and Defenses.** The security of 2G/3G text service is not without problems. The providers may suffer from

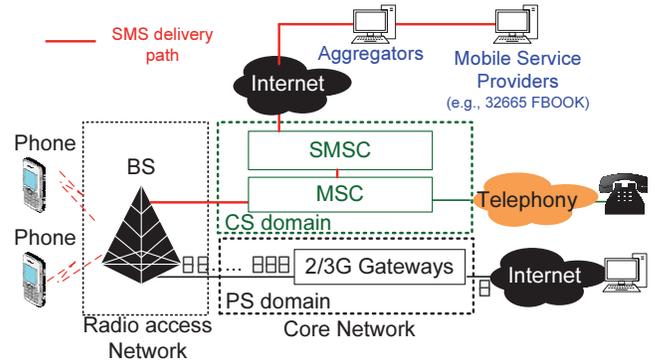

Figure 1. 2G/3G messaging service architecture, and mobile service providers with short code services.

unauthorized text delivery from phone-side malware [24], [25] and spoofed message [52] from Internet. Note that the spoofed message from mobile devices is forbidden in 2G/3G since there is no originating address header field for the text message being sent by SMS protocol design. More details will be given in §IV

Fortunately, most of these attacks towards mobile service providers have been effectively mitigated after years of active operations. First, for the phone-side malware, Android OS will automatically prompt users and ask for their confirmations while any application tries to send text messages to the short codes of mobile service providers. The details will be elaborated in §IV. Second, for the spoofed message from Internet (it is more likely done through the customers of the aggregators), more and more aggregators disallow their customers/users to send texts using phone numbers which do not belong to them (e.g., twilio, no spoofing [17]). Besides, the easiest way for mobile service providers is to explicitly ask their aggregators only forward the text messages from carrier networks. Note that we do not consider the scenario that malicious carriers compromise users' text messages here since it is not a very common case in practice.

### III. NEW SECURITY ISSUES WHEN TEXT SERVICE TURNS FROM 2G/3G CS INTO 4G PS

#### A. Text Service over PS Primer

Similar to the counterparts in 2G/3G, the text services in 4G LTE follow the similar reference model. The major difference comes from the underlying network infrastructure upgrade in 4G. Different from 2G/3G, 4G LTE is an IP-based mobile network. Its text service inevitably embraces the full packet-switched (PS) design. As shown in Figure 2, the text messages are no longer delivered through the circuit-switched domain. Instead, the messages are forwarded to the SMS gateways via 4G PS gateways, IMS (IP Multimedia Subsystem) and IP-SMS GW (IP-to-SMS Gateway) through SIP (Session Initiation Protocol) [40]. The underlying network technology evolves largely, however, most service providers do not update their mobile service in accordance with the 4G text service upgrade. However, we find that such defenses fail to fully protect mobile services. As we will show next, potential vulnerabilities are exposed to mobile services over 4G text service.

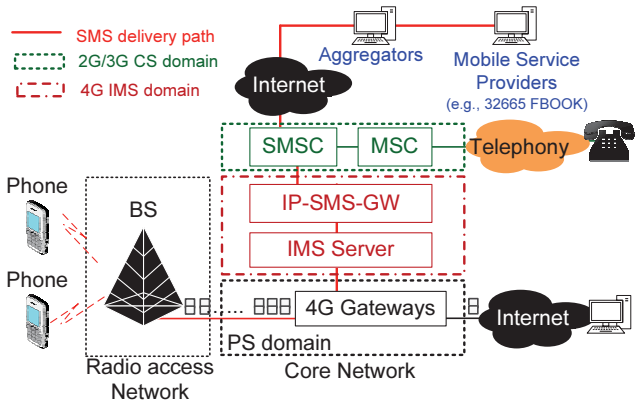

Figure 2. 4G text service architecture.

## B. Potential Vulnerabilities

Ultimately, text and data operate services in the same, connection-less IP network. However, this paradigm shift is double-edged, exposing mobile users, carrier and mobile service providers to unanticipated vulnerabilities.

In this paper, we look into three security aspects.

1) How do the phone-side security mechanisms accommodate the significant change in mobile networks? Are they still working fine or can be bypast by the attackers to launch a novel attack?
2) Are carriers ready to launch the text service over PS? Does the current operational network infrastructure well protect both mobile service providers and users from malicious attacks? Or, on the contrary, it help attackers to broaden the damages of attacks in the real world?
3) Does there exits any security loopholes of current service enrollment, use scenario or runtime authentication mechanisms which can be exploited by attackers in 4G LTE networks? Are the current mobile services still well protected as they cooperate with the 2G/3G text service?

Our study covers three major components of mobile services via text: mobile devices, carrier networks, and service providers. In the following sections, we disclose how the currently employed or newly developed mechanisms fail to harden mobile services against attacks and how they are exploited to abuse Facebook account(s) (§IV and § refsect:large-scale), donation attack (§VI) and spam lawsuit attack (§VI).

## C. Attack Model and Methodology

The attacker is a mobile user, whereas the victims can be the mobile service providers or other mobile users. The adversary mainly uses a commodity smartphone which does not have the rooted privilege to launch attacks. (S)he has no remote access, at least no privileged access to the victim phones. In all our attacks, we do not require any malware to be installed on the victims' phones. However, to avoid the tracking issue, attacker may distribute an unprivileged Android application (malware) to other mobile users and turn them into the helpers to launch attacks. Note that this application does not require any the permission to send or receive text messages. Besides, in all cases, we assume that the attacker has no control to the carrier network.

To validate vulnerabilities and attacks, we conduct experiments in two top-tier US carriers denoted as OP-I and OP-II, for the privacy concern. They together cover almost 50% of market share [7]. We use two Android phone models that support SMS over IMS domain: Samsung Galaxy S5 and LG G3, running Android 4.4.4 and 4.4.2, respectively. Note that only a few recent models support this feature, because it requires phone hardware and software upgrades. We focus on the Android OS but we believe that the identified issues are applicable to any other OS.

We bear in mind that some feasibility tests and attack evaluations might be detrimental to users, operators or mobile service providers. Therefore, we conduct this study in a responsible manner through two measures. First, we use only our own phones as the victims. Second, we purchase unlimited text plans for all tested phones. We seek to disclose new security vulnerabilities brought by SMS over IMS and effective attacks, but not to aggravate the damages. We do not claim that the attacks are not the most powerful ones to make damage or there exists no ways to detect them.

## IV. CASE STUDY: ABUSING A FACEBOOK ACCOUNT WITHOUT PASSWORD

In this section, we use Facebook as a case study to illustrate the vulnerabilities exposed to mobile service customers using SMS. We report a contactless attack where an adversary can remotely manipulate a victim's Facebook account without knowing his password. The attacker can update the victim's status, like a page/status, add friends, subscribe/follow/poke someone, on the victim's behalf without his/her consent. More threatening, the attack itself is not intricate. It is ready to launch by any individual with a commodity phone or device[6] which supports SMS service in 4G LTE networks. Note that our approach is also applied to Twitter as well. Due to space, we only present Facebook as a case study.

We find out that the vulnerabilities are rooted in multi-fold interactions among the service provider (here, Facebook), mobile device, carrier networks and even cellular standard organization (3GPP2). The service providers (*e.g.*, Facebook) leverages the mobile service (here, SMS) and mobile devices to offer convenient and friendly user experience. It counts on user authentication offered in cellular networks to protect their own service access. It thus takes authenticate-once-use-forever scheme where it merely authenticates users once, but allow them to access services forever. However, it is not secure while both carriers and mobile devices do not offer the guaranteed guard any more with their technical upgrade and evolution. In this case, when the cellular network technology is moving into 4G LTE, SMS is migrated from the conventional CS domain to the PS domain. Its evolution unfortunately exposes SMS to an open attack interface and thus allows a contactless impersonation. More surprisingly, we disclose that this problem is not caused by imprudent implementation bugs, but rooted in the problematic design regulated by the standard organization. The standard aims to offer flexibility (several options) to carriers to develop a better SMS service but some options are inherently lacking sufficient protection as they did before. This change eventually makes the whole chain collapse.

---
[6]It can be a tablet or a computer with cellular modem.

*A. Security Loopholes*

*1) Facebook, authenticate-once-use-forever:* The first vulnerability lies in Facebook's authentication scheme: Authenticate-Once-Use-Forever. In fact, in order to secure user's account (see the official page in Figure 3), Facebook encourages and even allures their customers to bind their accounts with their mobile phone numbers. The reminder of registering a phone number keeps prompting if the users does not yet. Consequently, a large number of users, especially technology-unsavvy ones, have done this. This seems solid for the following reasons. First, mobile phones are more personal devices and are almost carried by people all the time. This offers a always-online experience to Facebook accounts. You can receive friends' latest status updates or messages through text. Second, cellular networks adopt SIM-card-based user authentication, which is typically more secure than the password protection. A mobile phone number is associated with an authenticated user unless the SIM card is hijacked (not a commonplace scenario though possible [8]).

To register a phone number, Facebook requires 4-step procedures as illustrated in Figure 4. The user first sends a request to "Add a Phone" after (s)he logins website. Afterwards, the user texts the letter "F" to 32665 (FBOOK) from his/her phone and receives a one-time confirmation code issued by Facebook. Finally, the user enters the confirmation code received at the previous "Add a phone" webpage. By this way, Facebook builds a secure association between a mobile phone number and a Facebook account. This is ensured three cascading authenticated channels. The first is the session between a Facebook account and Facebook, which has been authenticated by the user password (marked in grey). The second is the SMS channel between a phone number and Facebook (via cellular network), guaranteed by SIM-card-based user authentication in cellular network (in a red box). The third is the one-time random code which stays valid in a limited time window (say, several minutes). Therefore, the code should be known only to the one who owns the phone number, not a third-party. As a matter of fact, such one-time two-way authentication via text has been widely used in other services, such as banking, e-commence, instant messaging, website registration, *etc.*, to verify the user identifiers.

The vulnerability is that the above authentication is one-time only. Afterwards, it does not run any extra authentication to ensure that the operation command comes from this registered phone number. The above registration by default turns on Facebook Text Service [4], which allows mobile users to update his status, add a friend, poke someone, like a page to name a few, through sending command messages to 32665 (FBOOK), *e.g.*, LIKE Facebook stories, ADD a Friend's email address or phone number. We conduct a 28-day experiment to perform six major functions provided by Facebook text service after the initial registration. We find that all functions can be performed without using any password or extra mechanism in the whole test cycle.

Such one-time message authentication fundamentally balances between security and usability. Despite seemingly sim-plistic, one-time message To invoke the password-based messaging authentication, the user should manually type in the message. Clearly, it is practically inconvenient to apply complex authentication/encryption operations upon each message, too difficult for manual input. Therefore, most applications (like Facebook) decide to fully trust the received messages at runtime, as well as the binding of the phone number, assuming that the mobile OS and the cellular infrastructure have ensured the authenticity of the message.

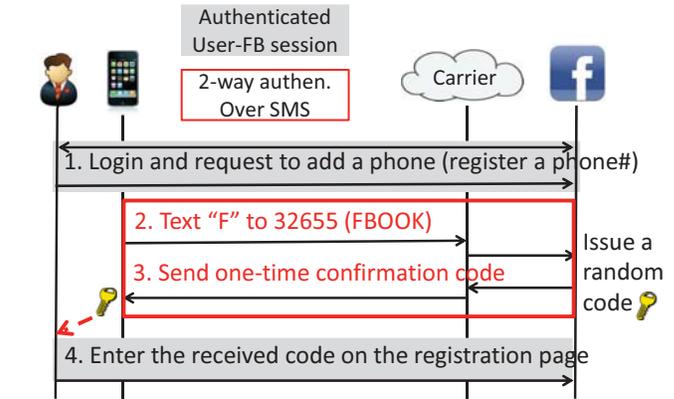

Figure 4. Procedure to add a phone and enable Facebook Text service.

*2) Conflicts between Cellular Standards, Negotiable Security Mechanisms:* For cellular networks, there exists two international telecommunication organizations: 3GPP funded by ETSI (European Telecommunications Standards Institute) and 3GPP2 funded by Qualcomm. They stipulate their own radio access technologies and core networks standards. General speaking, compared with complex cellular-specific protocols in 3GPP, 3PP2 is more open to offer carriers options (e.g., how to communicate with IMS server) in operations or utilize the techniques (e.g., Mobile IP) have been broadly used in Internet. We notice that there exists some differences/conflicts between two cellular standards might lead to security loopholes in 4G LTE.

**Security mechanisms[7] are negotiable in 3GPP2 but not in 3GPP.** 3GPP2 standard [21] offer five security options (TLS, DIGEST [40], IPSec-IKE, IPSec-MAIN, IPSec-3GPP) to carriers to determine how secure their 4G users to communicate with the IMS servers, whereas 3GPP leaves carriers no choices but IPSec-3GPP to establish secure communication between mobile users and IMS servers. While DIGEST is chosen (i.e., authenticate the mobile user while the SIP registration procedure is conducted towards IMS server), there is no network-layer or transport-layer security protection for the communication between mobile users and the IMS servers. It relies the underlying cellular-specific security mechanisms (e.g., confidentiality and integrity protection for the data delivered between mobile devices and base stations.

Our experiment results show that OP-I (3GPP2 carrier) adopts DIGEST in its 4G LTE networks as shown in Figure 5, whereas OP-II (3GPP2) uses IPSec-3GPP to encrypt/decrypt the SIP messages. We use *Wireshark* to capture packets delivered between mobile device and carrier networks. The mobile deceive receives a SIP 401 Unauthorized after it sends `SIP REGISTER` to the IMS server. In the `SIP 401 Unauthorized` message, the IMS server specify the method of user authentication to be DIGEST and provide parameters required (e.g., challenge, algorithm, nonce). The mobile device computes the answer to auth challenge and submits a new `SIP REGISTER` to the IMS server. Afterward, the mobile device receives a unencrypted `SIP 200 OK` from the IMS server. Note that this response and all messages after it shall be encrypted if IPSec is enabled (the keys for IPSec are computed during the digest authentication by cellular standards). It shows that OP-I deploys DIGEST instead of IPSec family into its 4G LTE networks.

---
[7]We focus on IMS security framework in this paper.

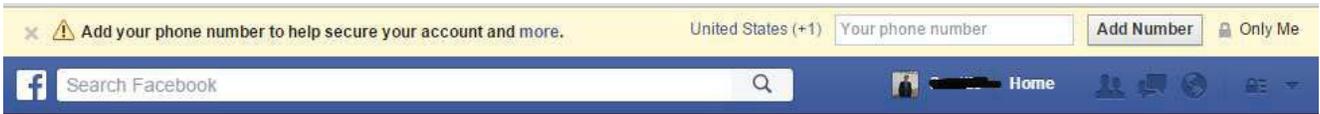

Figure 3. Facebook asks users to add their mobile phone numbers to secure their accounts.

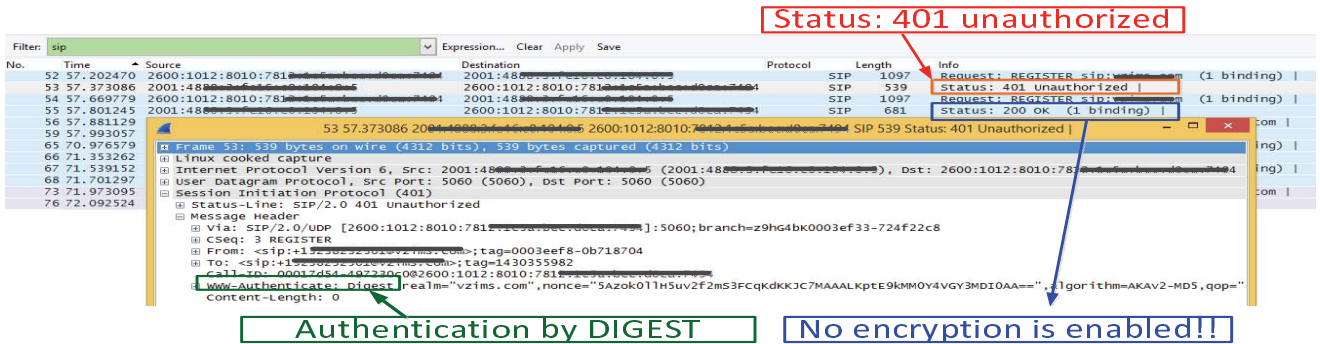

Figure 5. Wireshark trace: A US major carrier adopts DIGEST as the main security mechanism in its 4G LTE network

Carrier might trust that the underlying security mechanisms have provide sufficient protection and the disabling of IPSec can improve the data delivery performance to some extent. Nevertheless, it can be exploited to analyze the communication between mobile users and IMS servers to devise attacks. We further learn how to compose a valid `SIP MESSAGE` carries a SMS being sent and the way to send it to the proper IMS server. More details are given in Appendix A and B.

*3) Differences between Cellular Systems, Not Ready for Originating Address Newly Introduced:* To prevent malicious users from sending text with unauthorized phone number (i.e., spoofed text), cellular standards does not allow users to fill sender's phone number into the text being sent by its design of mobile text protocols (SM-TL [20]). There is no "originating address" header in the text message being sent in the legacy 2G and 3G systems. Nevertheless, the design principle is no longer available in the 4G LTE standards.

While a 4G user sends SMS, he/she need to specify the sender phone number into the `From` header of `SIP MESSAGE` which carries the SMS towards the IMS server as shown in Figure 6. We believe that it is a new threat/challenge to carriers since they do not need to concern the spoofed SMS from mobile devices in legacy systems. Once 4G LTE carriers do not notice it or deploy the necessary security mechanism (i.e., verify sender identifier in IMS server), it will be exploited to launch a new device-based spoofed SMS attack which differs from internet-based one disclosed in the early researches.

Unfortunately, our experiment results confirm this suspi-cion in OP-I. We conduct an experiment to identify what kinds of phone numbers an be spoofed. It works as follows. We prepare a OP-I mobile device whose phone number is XXX-YYY-2501, compose `SIP MESSAGE` carries SMS message by fill in the originating address to the phone number being verified and destination address to XXX-YYY-2501, and sent it the IMS server based on the approaches introduced in Appendix A and B. If our OP-I mobile device receives a SMS message comes from the phone number being verified, it means that the sender's phone number might be spoofable. We further construct the same `SIP MESSAGE` but use different destination address (e.g., another mobile phone number of OP-I or distinct carriers), send it out and examine if there exists any restriction to receive such spoofed SMS.

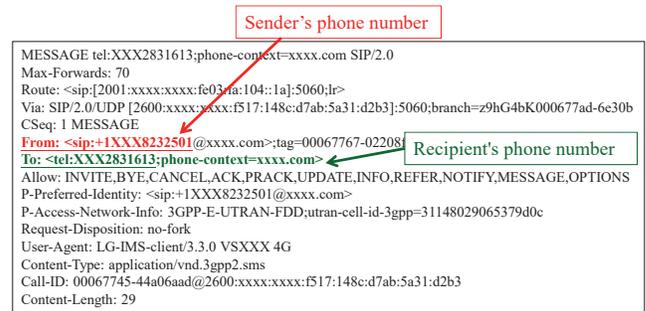

Figure 6. 4G LTE users are allowed to specify their phone numbers in `From` header in the `SIP MESSAGE` carried the SMS being sent while they communicate with IMS servers.

We verify the sender and recipient phone numbers belong to the top four largest US carriers including OP-I and OP-II. We find that only mobile phone number belongs to OP-I is spoofable in OP-I 4G LTE networks. It shows that OP-I indeed deployed the security check for the originating address newly added to 4G LTE. However, it is still far away the satisfaction. Second, there is no sign of restriction for users to receive the spoofed SMS sent from OP-I.

*B. Proof-of-Concept Attack*

We devise a proof-of-concept attack to show how attacker manipulate another user's Facebook account though Facebook

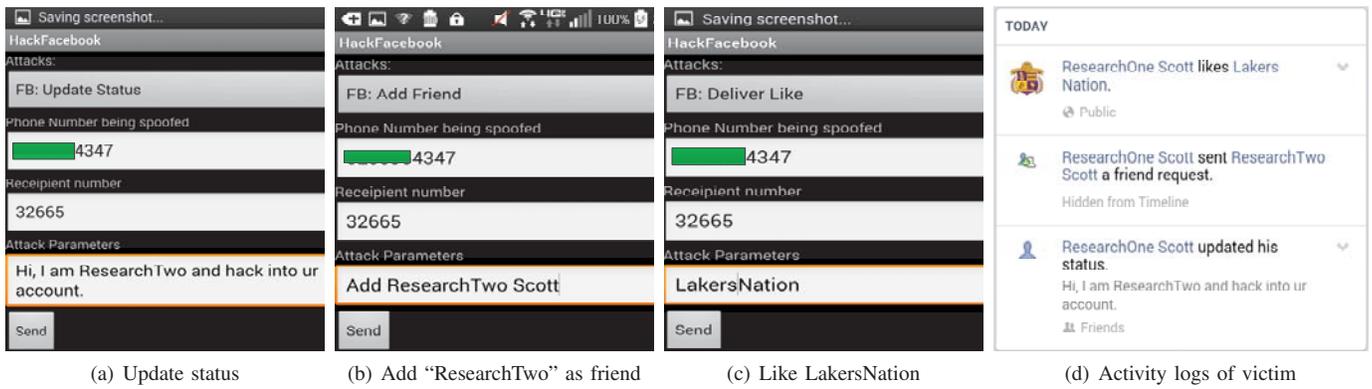

(a) Update status    (b) Add "ResearchTwo" as friend    (c) Like LakersNation    (d) Activity logs of victim

Figure 7. Launch *update-status*, *add-a-friend* and *like-a-page* Facebook attacks towards ResearchOne from ResearchTwo who is using the phone number 2501 (last four digits).

Text Service to update status, add a friend and like a page without his/her permission. We called them as *update-status*, *like-a-page* and *add-a-friend* attacks afterwards. Note that no malware is required to be installed on the victim's mobile phone, computer or any device can access the Facebook service.

*1) Design:* Clearly, the above loopholes can be exploited by attackers to proceed all functions of Facebook Text Service on behalf of victim. It works as follows. The adversary constructs the `SIP MESSAGE` carries SMS message based on the procedure in Appendix A with three modifications. First, victim's phone number is used as the originating address in all related SIP headers. Second, "32665" (i.e., FBOOK) is used for recipient phone number in both SIP headers and SMS headers. Third, we fill the specific commands (e.g., update-status: "Hi...", add a friend: "Add Bob", like a page: "Like Lakers Nation") into SMS content (`user data` shown in Figure 13) for Facebook Text Service. Then, attacker sends the `SIP MESSAGE` to the IMS server.

*2) Prototype and Evaluation:* We implement the *HackFacebook*, an Android application which only requires Internet permission to communicate with the IMS server, as shown in Figure 7(a). It can be launched on the adversary's own mobile device or Bots on Botnet (i.e., we can distribute *HackFacebook* as malware[8] to other mobile devices). To accommodate the various mobile devices with different Android versions (from 2.3.3 to 5.1.1), we are only using the socket APIs which are supported in all Android releases.

To evaluate the *HackFacebook*, we further create two Facebook accounts (ResearchOne and ResearchTwo[9]) and associate two OP-I mobile phone numbers with them (4347[10] is for ResearchOne, 2501 for ResearchTwo). We run *HackFacebook* on the mobile device using phone number 2501 to launch *update-status*, *add-a-friend* and *like-a-page* attacks towards the victim ResearchOne who is using phone number 4347. Figure 7 illustrates how those attacks are launched through

---

[8] For the malware version, the UI components shall be hidden and provide users with the certain network services, e.g., query the standard time. However, we don't discuss how to implement the malware version and prevent users from recognizing it since it is not the main focus of this work.

[9] Note that the naming of FB accounts does not present any information of authors.

[10] For privacy concern, we only show the last four digits of the mobile phone number here.

*HackFacebook* and the snapshot of activity logs of victims. It shows that the adversary can freely update status, add a friend (adds ResearchTwo as friend) or like a page (likes LakersNation) on behalf of victim without his/her permission.

*3) Variant, still works fine even sender spoofing of SMS is not allowed:* People may argue that our attack is devised on the fact that carriers do not perform strict security check for the originating address of SMS messages being sent. Once the security check is deployed, all issues we present above will be completely addressed. We are eager to clarify it here. As what we described earlier, the *HackFacebook* can be distributed as a malware and infects almost of all mobile devices to send SMS commands to Facebook Text Service since it does not require the root privilege. However, next question will be "*What is the difference between HackFacebook and malware-based text attacks if sender spoofing is forbidden in carrier network*"? There are uncounted malware developed to send SMS messages in background without user's consent. Is there anything new? We would like to answer this question in the following section.

*C. What's difference between current attack and early SMS attacks?*

We next highlight the differences between our attack models with previous two major SMS attacks.

**Phone-side malware text attacks** In recent years, the security research communities and public media [1], [2], [11], [55] have reported many malware, e.g., AnserverBot, HippoSMS, that send SMS messages to specific phone numbers in background. They do not intend to fake the SMS sender phone number but aim to launch the premium-rate SMS charging attack or leaking victim's personal information (e.g., location), to name a few. Most of them do not require the root privilege but SMS_SEND permission, and can infect significant mobile devices (e.g., 30,000 [1]). However, benefit from Google's research efforts, Android users are not threatened by the malware-based attacks towards mobile services, e.g., Facebook Text Service. It works as follows. While any Android application tries to send a SMS message to specific phone numbers or short codes (e.g., 32665, FBOOK), user will get a confirmation dialog popup which is controlled by Android OS as shown in Figure 8. Before user confirms it, no SMS message will be sent, even the SMS sending application has been granted with SMS_SEND permission. It effectively reduces the damages of the early phone-side malware text attacks.

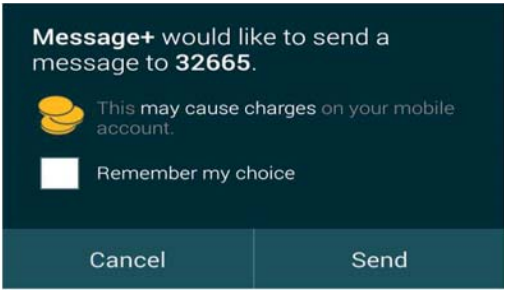

Figure 8. Android's security check on recipient's phone number

Different from the early phone-based malware text attacks, our attack can easily bypass the current Android's security check on recipient phone number. The root cause is that the security check is merely associated with Android text sending APIs (e.g., `SmsManager.sendTextMessage`). Once attacker is capable of delivering text messages via other approaches (e.g., by UDP socket APIs), the security mechanism is no longer effective.

**Network-based spoofed text attacks** There are several online text spoofing sources [13]–[15] providing their customers to send the spoofed text messages to any recipient. However, there are two main drawbacks to launch the network-based spoofed text attacks towards mobile services in practice. First, most of mobile service providers communicate with the mobile users by the unique short codes (e.g., 32665, more examples will be given in §VI). Only the authenticated aggregators can deliver text messages to the short code owners (i.e., mobile service providers here). It is not difficult for them to identify whether the text messages for mobile service providers are sent by wireless carriers or other sources (e.g., the customers of aggregators or content providers as shown in Figure 2), and further block all text messages sent by the non-carriers sources. Second, attackers are very easy to be traced back since those online text service providers including text spoofing websites do not provide the *free* text delivery service. They require users to setup the accounts and *pay* for the text delivery. From technical perspective, we do believe that the attackers might launch the spoofed text attacks towards mobile services and probably achieve the similar effects as our approach. However, it is difficult for them hide themselves due to those pre-established user accounts and the payment history.

Our attack differs from the network-based spoofed text attacks. First, the spoofed text messages are sent to mobile service providers through the legitimate wireless carriers and aggregators. The aggreagtors cannot protect the mobile service providers by blocking all text messages sent by non-carrier sources. Second, our approach does not require attackers to subscribe any voice/data/messaging plan from carriers or setup any account before launching attacks. Since *HackFacebook* can be distributed as a malware and infect the mobile devices to send the spoofed text, as well as bypass the security check mechanism deployed at mobile devices. The dual characteristics make the adversaries hard to be captured or detected. Note that we do not claim that there is absolutely no way to discover the attackers. However, our approach provides the attackers with the better camouflage than network-based attacks.

*D. Lessons Learnt*

There are four lessons we learnt. First, the runtime authentication is essential to the mobile service providers which provides the non-query-only services (e.g., update status, like page, etc). Two, the security is not negotiable regardless of any concerns. Mobile network standards should not offer an insecure communication mechanism (i.e., disable TLS/IPSec) to carriers. Third, the phone-side security mechanisms provided by OS (Android) cannot always capture the changes of underlying networking technology. We do need a more generic and flexible security framework for mobile device and carrier networks. Fourth, carriers should not be too rush to launch a service, especially for ones have incompatible issues with the early deployment, and create new security loopholes (e.g., incomplete security check on the sender identifier of 4G text message).

V. LARGE-SCALE ATTACK TOWARDS FACEBOOK USERS

In this section, we introduce how to scale the individual Facebook attack to large-scale attacks. We devise two novel attacks: (1) Free Million(s) Facebook Likes and (2) Large-Scale User Privacy Leakage, implement the proof-of-concept prototype and evaluate it.

*A. Free Million(s) Facebook Likes*

This attack aims to collect as many as Facebook Likes we can from the *distinct* real users on a given Facebook page. The number of Likes on the Facebook page not only indicates how popular it is, but also acts as an important factor affecting whether online advertising vendors post ads on the page (owner of Facebook page will get paid) and pay for it. According to [18], the real value of "Like" has been risen to $174 in 2014 (from $136 in 2010). Buying friends through social media is a common business practice which has been reported in [19]. There exists several emerging Like-Promotion companies (*e.g.*, GetPaidforLikes.com or YouLikeCash.com), which help owners of Facebook pages to increase the number of Likes. Attackers can launch this attack to collect significant Likes on his/her favorite Facebook pages or the advertisers' pages.

For the advertising industry, the current business model works as follows. First, Like-Promotion companies recruit Facebook users to install their own Facebook applications into participants' Facebook accounts. Then the application will automatically commit the Likes to the advertisers' pages by Like-Promotion companies' instructions. Each participant will earn around 1-5 cents per Like. The maximum number of Likes they can offer to the advertisers rely on the number of Facebook participants. Currently, there are 75K-460K users in GetPaidforLikes.com and YouLikeCash.com, which represent a business scale of $3,750-$23,000 per Facebook page (calculated based on the cost of 5-cents/Like the advertisers pay). If attacker can manipulate 100 millions Facebook accounts based on the attack model introduced in §IV and deliver their Likes to specific Facebook page, the benefit in dollars will be 5 million per page basis ($0.05×100 million Likes). The attackers can

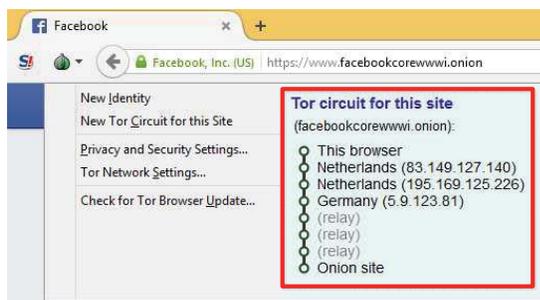

Figure 9. For anonymity, access Facebook by ToR-Browser. Note that we do not use it in Netherlands.

run online Like-Promotion companies outside US (avoiding trace-back via cash flow).

**Different from using fake Facebook accounts.** People may think that we can achieve the same effects (i.e, collect million Facebook likes) by creating significant fake Facebook accounts to like specific pages. Why should we use your approach? Yours is even more complicated. This argument sounds reasonable, however, the research community has been aware to those issues and proposed many mechanisms [26], [43], [49], [54] to detect fake accounts or spammers in social networks. Differ from them, our attack model obtains free Facebook likes through real Facebook accounts instead of fake ones. It brought two benefits. First, the current detection mechanisms of fake accounts are not designed for real user accounts. Second, the advertisers are more willing to pay for likes if they come from real users instead of fake ones.

### B. Leak Million(s) Users' Privacy

This attack aims to obtain the information about victims' names, jobs, friends, family members, schools, photo, etc., through *add-a-friend* attack towards Facebook Text Service. We first describe how it is operated since it slightly differs from the previous Facebook Like attack, and then show how attackers gain from it.

This attack works as follows. First, the adversaries create several Facebook accounts with the fabricated personal information (e.g., use temporary email addresses, http://10minutemail.com). Second, the attacker adds his/her pre-created Facebook account to the victims' friends by the methodology introduced in §IV. Finally, Facebook sends a add-a-friend notification to the adversary's Facebook accounts and requests for the approval. After approving this request, the attacker becomes a friend of victim and see all information (*e.g.*, phone number, school, job, living area, contact list, pictures, to name a few) that victim shares with his/her friends. By this approach, the adversary can associate the mobile phone number with victim's real profile and sell those personal information. Besides, to prevent Facebook from tracking attackers while they access their Facebook accounts (i.e., approve add-a-friend request), they can use ToR-Browser [16] to access Facebook as shown in Figure 9 for anonymity purpose. Though some researches challenge the anonymity of ToR networks. However, their prerequisites are not easily satisfied in practice (e.g., require NetFlow traces from the intermediate routers in ToR [27]). We thus believe that ToR still provide the attackers with the anonymity to a large extent in nowadays.

We next discuss another incentive model of this attack.

**Celebrity Effects.** The user privacy leakage itself might not bring the instant benefits to the adversary at the first sight. However, through the large-scale screening process above, there exists a decent chance to discover some popular celebrities on Facebook. If attackers successfully find some superstars, e.g., Cristiano Ronaldo (105M fans), Shakira (102M fans), Vin Diesel (94M fans), Eminem (92M fans), Will Smith (74M fans), the impact will be wide or even global. If we quantify it in dollar, a recent report claims that Oscar-winner Jared Leto garnered an astounding $13,000 for one sponsored social media post (single post!!). The current number of fans of Jared Leto on Facebook is 3.9M until August 2015. We believe that security loopholes identified deserve more immediate research efforts, since they indeed provide the attackers with the sufficient incentive to launch attacks in practice.

### C. Proof-of-Concept Attack

*1) Design and Prototype:* The design and implementation of the two large-scale attacks are similar to the individual *like-a-page* and *add-a-friend* attacks are introduced in §IV-B, except for two differences. First, for both of the million Facebook likes and user privacy leakage attacks, we have to prepare the a set of spoofable mobile phone numbers (candidate victims, OP-I users) before launching attacks. The *HackFacebook* selects a number from the victim candidate set, proceeds the individual attack and repeat the procedure until all numbers are explored.

Second, since not all Facebook users will share their mobile phone numbers with friends, we thus slightly revise the *add-a-friend* attack as follows. After *HackFacebook* sends the Facebook text message carries add-a-friend command to 32665 (FBOOK), it immediately sends another Facebook text message to update the victim's status by specifying his/her mobile phone number (i.e., the originating address of Facebook text messages being sent). After the adversary accepts the victim's add-a-friend request, the attacker can obtain victim's mobile phone number from his/her latest status update.

*2) Evaluation:* First, we need to clarify that our proof-of-concept prototype does not aim to produce real world damage but provide the feasibility studies. Thus, we here mainly evaluate the capability of *HackFacebook* to send a number of SMS messages at mobile device. To avoid legal issue, we conduct the evaluation experiment as follow. We use *HackFacebook* to deliver Facebook Text commands (i.e., for add-a-friend and like-a-page attacks) to ourselves. Once the Facebook text command is received, we send another Facebook text command to ourselves. Then we measure how many SMS messages *HackFacebook* can transmit and receive in 30 minutes. For performance comparison, we also conduct the same experiment through the default messaging application of Android. The results are plotted in Figure 10.

We have three findings. First, there is cap for the number of SMS messages sent by Android messaging app within 30 minutes, says 30 messages. After the number of messages being sent exceeds the threshold, Android OS keeps warning users and hold the message deliver until user explicitly approve it (press Allow button). We believe that it shall be Android's security mechanism (rate control) to protect users from the messages attacks launched by malware. Second, *HackFacebook* can suppress the warning dialog, and deliver up to 2459 messages in 30 minutes in the experiment conducted before April 2015. Third, we notice that the OP-I seems to deploy a network-based rate control which limit the number of SMS messages being sent in April 2015. However, *HackFacebook* is still capable of delivering up to 1002 SMS messages in 30 minutes which is 33x time faster than Android. Though there might exist some caps which we cannot discover in current experiment settings, the adversaries can mitigate the limitations by increasing the number of mobile devices infected to achieve the same effects they need.

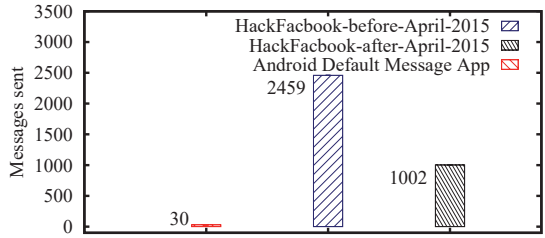

Figure 10. The number of messages are successfully sent and received in 30 minutes.

**Short-lived impacts, no strong incentives for attackers?** Some people argue that impacts of two attacks are short-lived. Once the victims discover that their Facebook accounts are abused, they will cancel the Like, unfriend the attackers, and report this issue to Facebook. Then the adversary gains nothing from these attacks in the long term. We need to clarify that our research does not aim propose the attack models for attackers to live on. Instead, we focus on the feasibility study, and real-world impacts analysis, and technical challenges to capture real attackers. How to guide the attackers make a big profit by them in a short time window (couple days or weeks) is not our major concern in this work. Even there are several options for them to make quick money (e.g., buy put option of Facebook before attacks, win the bet (bet if I can make Well Smith to add you as friend)).

### D. Lessons Learnt

As what we learnt earlier, the phone-side security mechanisms (e.g., text sending rate control) provided by OS is hard to take the immediate actions as underlying networking technology evolves. A more open and device-network cooperation security framework should be introduced to together defend against new security threats while networking technology changes. Moreover, we believe that carriers should share the blame due to its loose regulation for text message delivery. This issue can be exploited to launch the phone-based large-scale text message attacks.

## VI. STUDIES ON MORE MOBILE SERVICES

In this section, we conduct a medium-scale study on the 64 mobile services from 27 distinct industries plotted in Table I. By thoroughly analyzing their service models, use scenarios, enrollment and runtime authentication mechanisms and , we discover several security loopholes of them, which are vulnerable to the security threats we presented in §IV and V. We devise two large-scale attacks: (1) fulfill billion(s) dollar donation attack and spam lawsuit attack to estimate the real world impacts of those vulnerabilities. Finally, we summarize the lesson learnt.

### A. Service model, use scenarios, enrollment mechanisms and runtime authentication

**Service model and use scenarios.** Among 64 mobile services, there are two main service models: (1) *request-response* and (2) *notification/alerts subscription*. The service model of the former is that users first send the request text to mobile service provider. The service server further process the request and respond users with the results (it is optional, depends on the mobile services). The former is usually adopted by mobile banking, social networking, mobile giving (donate money from phone), flight status query and etc. The latter is that users subscribe the certain notifications (e.g., special offer, flayer, news update, delivery status of goods, friends updates their Facebook/Twitter) or alerts (credit cart payment due or abnormal transactions).

**Enrollment mechanisms.** We observe the six enrolment mechanisms summarized in Table II. *One-Step* allows users to enroll the mobile service through website without username and password (e.g., sign up phone number on http://m.target.com/sms/baby/form to receive Target Baby's promotions) or SMS (e.g., text "JOIN" to 555444 to subscribe Staple's special offer and update). It is the most convenient and popular way (23/64) to subscribe the mobile service. *Two-Step* requires users to login the mobile service account on websites and enter the mobile phone numbers on website. This approach is not common (2/64) but adopted by two major US banks including US Bank and Discover. *Three-Step-Simple* is similar to *One-Step*. The main difference is that the after users enroll the service, they will receive the confirmation text from the service providers, and respond them with the *fixed* responses (e.g, YES, Y) to complete the subscription. The simple and *fixed* reply message is the main reason we call it as *Three-Step-Simple*. The *Four-Step-Simple* and *Four-Step-AuthCode* require users to login the mobile service on websites, enter phone numbers, wait for the confirmation text messages. The main difference is that the confirmation text message in *Four-Step-AuthCode* will carry a authentication code. User has to fill it in on the mobile service website, whereas, *Four-Step-Simple* merely asks users to reply the *fixed* text messages (e.g., YES). The advocates of the two approaches are mobile banking, social networking and online payment service providers which concerns user security and privacy more.

The last one is *Always-On*, it means that mobile users do not need to subscribe the mobile services since carriers have done it on behalf of users. Currently, we only observe it on the mobile giving service which donate few money ($5-$10) to some non-profit organizations, e.g., Red Cross or Food Bank. Users don't need to provide the charities with your checks, credit card number or banking account information. Take Red Cross as example, user just texts REDCROSS to 90999 to give $10 to American Red Cross and his/her carrier will charge it in the month bill or the balance of pre-paid card.

**Runtime authentication.** Last but not least, to our surprise, after users finished mobile service subscription, up to 50 mobile services do not request for any runtime authentication (i.e., ask users for passwords or secret codes before executing the requests) while users access the mobile services. For the notification-subscription mobile services, it might be acceptable since it is a convenient way to users to receive those notifications. However, for the request-response mobile services, e.g., Facebook or Twitter, this approach can be exploited to abuse their accounts as what we show in §IV and V. Though the remaining 14 mobile services deploy the runtime authentication, however, it cannot be considered the strong one. It works as follows. On receipt of the request, the mobile service server simply responds users with "Are you sure to do XXX. If yes, please reply YES" as a simple runtime authentication.

| Enrollment Method | Service Model | | | | | | Total |
|---|---|---|---|---|---|---|---|
| | Request-Response | | | Notification-Subscription | | | |
| | Web | SMS | Carrier | Web | SMS | Carrier | |
| One-Step | 0 | 0 | 0 | 5 | 18 | 0 | 23 |
| Two-Step | 2 | 0 | 0 | 0 | 0 | 0 | 2 |
| Three-Step-Simple | 0 | 0 | 0 | 5 | 9 | 0 | 14 |
| Four-Step-Simple | 2 | 0 | 0 | 1 | 0 | 0 | 3 |
| Four-Step-AuthCode | 6 | 0 | 0 | 0 | 0 | 0 | 6 |
| Always On | 0 | 0 | 14 | 0 | 0 | 0 | 14 |

Table II. SIX ENROLLMENT MECHANISMS FOR MOBILE SERVICES.

### B. Security Loopholes

We discover two security loopholes from the design of current mobile services.

**Loose enrollment mechanisms of the notification-subscription services.** 37 of 38 the notification-subscription services adopt the *One-Step* and *Three-Step-Simple* as their subscription mechanisms. However, they cannot tell whether the subscriptions are coming from the real users or attackers. Our tool *hackFacebook* can be easily revised to subscribe those notification services on behalf of the victims. For the services using *One-Step* subscription approach, we directly texts the subscription command to the mobile service provider. For the ones using *Three-Step-Simple*, we first texts the subscription command to the service provider, wait for 5 seconds, and texts the pre-defined *fixed* reply message (e.g., YES, Y, GO) to the provider. Note that the attacker is not required to receive the confirmation text used in the *Three-Step-Simple* and finish the service subscription.

**Always-On service enrollment and weak runtime authentication used by a number of money-involvement services.** To encourage people to support the charities or non-profit organization, the most of carriers enroll the mobile giving services on behalf of their customers. For example, there are forty carriers (Verizon, AT&T, Sprint, T-Mobile, etc) participating in American Red Cross's Text Message Donations program. After the enrollment, their customers can easily donate money to those non-profit organizations. However, is the security mechanism of mobile giving services revised accordingly with the evolution of mobile text services in 4G LTE networks? Unfortunately, our studies conduct a negative answer. Most charities only provide their donators with the weak runtime authentication (i.e., To confirm your $10 donation to XXX rely with YES or billing zipcode). For an attacker who can bypass all security checks at mobile devices and carrier networks, the always-on mobile giving service enrollment with a simple runtime authentication mechanism can be exploited to launch a financial attack in US.

### C. Donation Attack

This attack aims to a small-scale financial disorder and crisis. The adversary selects one representative charity supports text messaging donation [6], and launch a large-scale donations on behalf of victims. We revise the *hackFacebook* used in large-scale attack. Use American Red Cross as example, it works as follows. For each victim, it sends two text messages to the 90999. The former is the text carries the donate code (i.e., REDCROSS). The latter is the text with "YES" (the *simple* and fixed the donation confirmation text). The time interval between the two texts is configured to $5^{11}$ seconds. During the waiting period, it can launch attack towards another victim (i.e., conduct attack in pipeline). By this approach, for each victim, the adversary can donate $10 on behalf of him/her. If attackers launch a large-scale donation, it is projected to make the financial crisis. The following customers' complaining, refunding, news report might cause a financial disorder and small crisis in US.

### D. Spam Lawsuit Attack

This attack is devised to launch attacks towards those notification-subscription mobile service providers (e.g., Walmart, HomeDepot, Macy, Costco, etc). Most companies target on the end-consumer market usually allow customers to enroll their special offer, coupon, flyer or update and receive them by text. Those companies are the *final* victims of this attack. It works as follows. We subscribe those notification services on behalf f victims by our tool since those mobile service providers merely use *One-Step* and *Three-Step-Simple* subscription mechanisms. Afterwards, those companies will keep sending the promotion update texts to the victims. Those texts will be considered as spam message. In US, every user is well protected by CAN-SPAM (Controlling the Assault of Non-Solicited Pornography And Marketing Act) and TCPA (Telephone Consumer Protection Act) from the spamming messages. The lawsuit for spam message is not rare in US. For example, Papa John's faced $250 million lawsuit over SMS spam in 2012.

### E. Lessons Learnt

There are two lessons learnt. First, for the notification-subscription mobile services, they should pay more attention to the subscription mechanisms. Otherwise, it will be a double-edged sword. The more convenient it is, the more likely the mobile service providers face the spam lawsuit. Second, for the request-response mobile services, all non-query services (e.g., update status on Facebook, donate money, etc) should not be offered without a solid runtime authentication mechanism. If not, the mobile services are more likely to be abused by the adversaries.

## VII. REMEDIES

In this section, we propose defense measures to protect the mobile service providers and mobile users as shown in Figure 11. Our solution also seeks to be 3GPP/3GPP2 standard and industry (short code service) compatible, thus facilitating fast deployment.

---

[11]It is configurable. In our experiments, 95% text delivery is finished in less 5 seconds.

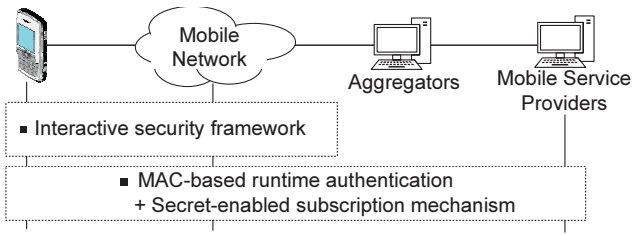

Figure 11. Overview of remedies to protect mobile services and users.

## A. Fulfill Interactive Security Framework

We suggest a interactive security framework shall be deployed between mobile device and mobile networks. From §IV and §V, we observe that Google indeed does its best to protect mobile users from malicious attacks (e.g., request for approval when there exists a SMS message being sent to specific phone number or warning dialog while a number of SMS messages are about to be sent). We appreciate their contributions to provide users with the secure SMS message service. However, the phone-based approach might not be best solution which can be easily bypass since they cannot always take immediate actions in accordance with the change of the underlying networking technology.

Thus, we propose an interactive security framework which provides a generic request-for-user-approval mechanism. [41] also propose similar idea, however, we try to make it more generic for all use scenarios. It works as follows. While carrier networks receives a special or attention needed request (e.g., send a SMS message to a premium-rate number or send a number of SMS messages in a short time), it replies the mobile user with a specific response code which requires user's approval. To be compatible with current mobile standards (all network services are provided through SIP in 4G LTE), we propose to add a new SIP response code, `SIP 440 User Approval Required`. It carries the same content of `SIP 401 Unauthorized` [40] which requests for user authentication, but a specific SIP header specify the information (e.g., "Sending a message to 32665 will produce extra cost. Do you agree with it?") being displayed to user. The mobile device treats the `SIP 440 User Approval Required` same as `SIP 401 Unauthorized`, and calculate the answers to the challenges. The only difference is that the mobile device only replies it after the user approve it.

## B. A Lightweight MAC-based Runtime Authentication

In recent years, both of academia and industry proposes several SMS security frameworks (SMSSec [34], PK-SIM [42], EasySMS [42]) or the end-to-end SMS encryption (Pushbullet [9]) to provide secure SMS communication between individuals and mobile service providers. For the infrastructure-based solutions, carriers need to modify current network architecture to support it accordingly. For end-to-end SMS encryption, it requires users to install the specific SMS messaging applications for the message encryption and decryption (e.g., Pushbullet). If so, why not use OTT messaging applications (e.g, Whatsapp)? Is there any better way?

In fact, the message encryption which most prior researches adopted is one of various methods to authenticate the message delivered between mobile service providers and mobile users. The most important thing for mobile service industry is to guarantee the authenticity of each message received. We revisit the current mobile service architecture in 4G LTE (Figure 2). We find that there is almost no ways for mobile users to deliver the spoofed SMS message (for 2G/3G users, it is forbidden by mobile network standards; for 4G users, it can be easily blocked at the core network). Besides, the delivery of SMS message in cellular network is confidentially and integrally protected. It implies that the delivery of SMS within carrier networks is still considered secure to a large extent (we do not consider that malicious carriers modify users' SMS messages, since it is not very common in practice).

Going down this direction, the possible attackers should be the aggregators, content providers or their customers who create spoofed SMS messages or modify the messages sent from wireless carriers to launch the occasional attacks (otherwise, they will be easily discovered and traced back). For those small-scale attacks, we believe that the lightwight message authentication code (MAC) mechanism with the symmetric secret key shared by the mobile users and mobile service providers can well defend against them. Though the MAC takes 16-20 bytes (use cryptographic hash functions: MD5 or SHA-1) from single SMS message (can carries 140 bytes data), it occupies around the 14% space which is still considered accepted for the better runtime authentication. To facilitate the deployment, we suggest to implement it in the default messaging application in Android. For the message which fail to pass the verification of MAC, it shows a warning icon which presents that it is a unauthenticated message or your mobile service provider does not support runtime authentication.

**Secret-based subscription mechanism** To support MAC-based runtime authentication, the prerequisite is to generate the symmetric secret key between users and service providers. It is not difficult for the mobile services which request for the setup of user accounts prior to usage (e.g., mobile banking, social networking, etc). The shared key can be the user' password. For other mobile services (e.g., Walmart's promotion service), they do not ask users to create the accounts during subscription for convenience. However, we believe that they still can generate a easy-remember secret code (e.g., WXM889) or allow users to define their own secret codes during the service subscription. Beside, to prevent the malware from intercepting the secret code through text, we suggest that it should not be exchanged through text but via websites enable SSL/TLS secure connections. Afterwards, all messages exchanged between mobile service providers and mobile users are appended the corresponding MAC.

## VIII. RELATED WORK

The mobile service security and privacy have been actively studied in recent years. Diverse threats are identified from mobile services, including permission bypass [37], [53], fraudulent money transfer with online banking [36], [39], faked social network follower market [44]. The general vulnerabilities include the coarse-grained Android permission control [23], [32], [33], misuse of Android's cryptographic APIs [29], [45], account hijack [46], side-channel privacy leakage [28], *etc.*. To defend malware, both static analysis [30], [35], [51] and policy-driven approach [50] are explored. Compared with these work, the vulnerabilities we discussed here are specific to the mobile text service.

For the traditional circuit-switched and signaling based text service, these early research efforts indeed help defend threats of user privacy leakage in the message [38], faking the message-based commands to the device [10], spamming the voice with signaling-based message [31], [48], malware/virus embedded in the message [24], [25], to name a few. In this work, we demonstrate that with recent deployment of packet-switched 4G LTE text service, these efforts are insufficient to protect the mobile services.

IX. CONCLUSION

Mobile service over text messages is still one of the most popular approaches by many service providers. By taking advantage of readily deployed text service infrastructure, this approach not only offers ubiquitous mobile services to all mobile users, but also benefits from the built-in security protections in cellular network. After years of operations and evolutions in 2G/3G network, this approach is proven to be effective in guaranteeing content integrity, origin authenticity and accountability.

In this work, we show that these guarantees could be violated in the 4G LTE context. The novel text service, as well as security shields deemed effective for 2G/3G text service, exposes new vulnerabilities in 4G networks. Large-scale attacks can be launched in diverse mobile services, including account-abuse, billion-dollar donation, spam lawsuit attacks, to name a few. The causes of these vulnerabilities are diversified. The mobile devices (OS) and mobile service providers do not update their security functions in accordance with the carrier infrastructure upgrade. For carriers, they indeed deploy new security functions while the text service is turned from 2G/3G CS into 4G PS domain. However, their incomplete security functions and imprudent operations (i.e., disable TLS/IPSec) further expose vulnerabilities to attackers. To defend against these threats, we propose remedies that are readily deployable and compatible with mobile network standards.

Three lessons can be summarized from these vulnerabilities and attacks. First, phone-side security mechanisms cannot accommodate the speedy and unpredictable the changes of underlying networking technology. We believe that a general, interactive device-network-cooperation based security framework can offer more flexibilities and defense measures to users and carriers to defend against the new attacks in the future. Second, the mobile network standards (3GPP/3GPP2) should not provide insecure communication options (*e.g.* disable TLS/IPSec) to carriers. Such seemingly flexible configurations may incur serve security concerns to not only the carrier network, but also mobile services on top of it. Last but not least, the runtime authentication is critical to guarantee per-message integrity, authentic and accountability. Mobile service providers should not expect to protect every content message with one-time authentication at registration phase.

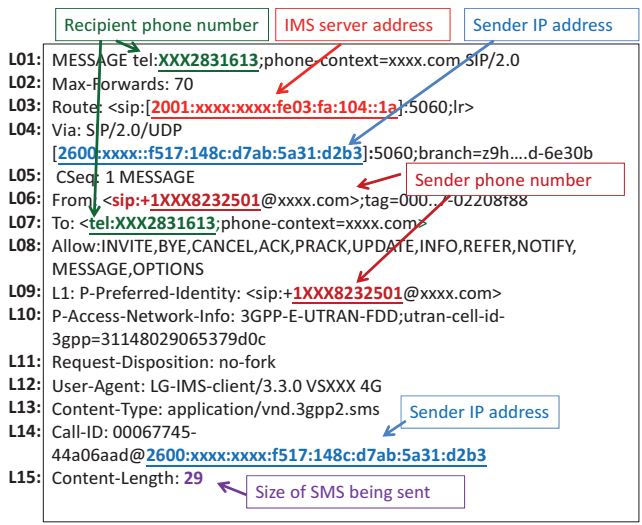

Figure 12. Message headers of `SIP MESSAGE` carries a SMS being sent.

APPENDIX

In Appendix, we introduce how to construct and submit a `SIP MESSAGE` carries SMS message in OP-I.

### A. Construct a `SIP MESSAGE` carries SMS message

To construct it, we need to prepare two parts: message headers and message body. In the following, we describe how to construct each of them on the mobile device.

**Message headers** Figure 12 illustrates the headers of a `SIP MESSAGE` contains an SMS message. We summarize that the eight header fields shall be filled accordingly by mobile device's realtime profile: Request-Line (Line 1, recipient's phone number), Route (Line 3, IMS server address), Via (Line 4, sender's IP address), From (Line 6, sender's phone number), To (Line 7, recipient's phone number), P-Preferred-Identify (Line 9, sender's phone number), Call-ID (Line 14, sender's IP address), Content-Length (Line 15, the size of the message body). For other headers, their values will not affect whether the `SIP MESSAGE` is accepted or rejected (i.e., we can apply same values to different mobile devices).

The most important step to construct message headers is to obtain the IP address of the IMS server since it is not static and open to public. We retrieve it from the routing table by executing a Linux command "ip -6 route". Note that this command does not require root privilege. On all of our test phones, there exist two network interfaces: `rmnet0` and `rmnet1` for the packets of the Internet data service and the IMS services (e.g., SMS delivery or VoLTE (Voice over LTE, akin to VoIP)), respectively. The IMS server shall be associated with the network interface reserved for IMS services. However, we found that the IMS network interface may be `rmnet0` or `rmnet1` in different phone models, we thus need to differentiate these two. We find that carriers alway associate the "Default" routing rule with the network interface assigned to Internet data service (e.g., default via fe80:xxx::5dc8 dev `rmnet0` metric 1024). Thus, we infer which network interface is reserved for the IMS services and obtain the IMS server address from its routing rule (e.g., "2001:xxxx:y:fe03:fa:104:0:5 via ... dev `rmnet1`").

**Message body** Figure 13 presents a CDMA SMS message [22] (for 3GPP2 carriers). There are two fields being filled accordingly: destination address and user data (the content of SMS). Android OS provides a set of classes (e.g., CdmaSmsAddress, BearerData) in the library, *ITelephony* [3] to construct the message.

### B. Send a `SIP MESSAGE` carries SMS message

We analyze the SIP signaling exchange between mobile devices and IMS servers, and observe a fixed communication pattern. All IMS servers are waiting for `SIP MESSAGE` signallings from mobile devices by listening to the UDP port number 5060. Besides, we notice that there is no strong binding between the SIP session and the underlying transport layer. Specifically, IMS servers will process `SIP MESSAGE` signallings from the mobile devices regardless of what the UDP source port numbers they uses during the user authentication. We next simply opens a UDP socket, specify the destination address and port number to be IMS server address and 5060, respectively, and send the `SIP MESSAGE` out.

Figure 13. Message body of `SIP MESSAGE` carries a SMS being sent.